\documentclass[prc,twocolumn,showpacs,preprintnumbers,amsmath,amssymb,superscriptaddress]{revtex4}

\newcommand{\nuc}[2]{$^{#1}$#2}
\newcommand{\nucm}[2]{$^{#1}$#2$^m$}

\newcommand{\qec}{$Q_\text{EC}$}

\def\F{{\cal F}}

\usepackage{graphicx}% Include figure files
\usepackage{dcolumn}% Align table columns on decimal point
\usepackage{bm}% bold math
\usepackage[latin1]{inputenc}

\begin{document}

\preprint{APS/123-QED}

\title{\qec-values of the Superallowed $\beta$-Emitters \nuc{10}{C}, \nuc{34}{Ar}, \nuc{38}{Ca} and \nuc{46}{V} }
% Force line breaks with \\

\author{T.~Eronen} \email{tommi.eronen@phys.jyu.fi}
\author{D.~Gorelov}
\author{J. Hakala}
\affiliation{Department of Physics, P.O. Box 35 (YFL), FI-40014 University of Jyväskylä, Finland}
\author{J.\,C.~Hardy}
\affiliation{Cyclotron Institute, Texas A\&M University, College Station, Texas 77843, USA}
\author{A.~Jokinen}
\author{A.~Kankainen}
\author{I.\,D.~Moore}
\author{H.~Penttil\"{a}}
\author{M.~Reponen}
\author{J.~Rissanen}
\author{A.~Saastamoinen}
\author{J.~\"{A}yst\"{o}}
\affiliation{Department of Physics, P.O. Box 35 (YFL), FI-40014 University of Jyväskylä, Finland}

\date{\today}

\begin{abstract}
The \qec{} values of the superallowed $\beta^+$-emitters \nuc{10}{C}, \nuc{34}{Ar}, \nuc{38}{Ca} and \nuc{46}{V}
have been measured with a Penning-trap mass spectrometer to be 3648.12(8), 6061.83(8), 6612.12(7) and 7052.44(10)
keV, respectively.  All four values are substantially improved in precision over previous results.
\end{abstract}


\maketitle

\section{INTRODUCTION}

Superallowed $0^{+} \rightarrow 0^{+}$ $\beta$ decay between $T=1$ nuclear analog states plays an important role in
several fundamental tests of the three-generation Standard Model.  It tests the Conservation of the Vector Current (CVC),
probes for the presence of scalar currents, and is a key contributor to the most demanding currently available test of
the unitarity of the Cabibbo-Kobayashi-Maskawa (CKM) matrix \cite{Towner2010a}.  For these and other reasons, it has been a
subject of continuous and often intense study for six decades.  The most important features of these superallowed transitions, 
and the ones that make them so attractive, are that their measured $ft$ values are nearly independent of nuclear-structure
ambiguities and that they depend uniquely on the vector (and scalar, if it exists) part of the weak interaction.

To date, the measured $ft$ values for transitions from ten different nuclei are known to $\sim$0.1\% precision, and three
more are known to between 0.1\% and 0.3\%.  An analysis of these $ft$ values \cite{Hardy2009} recently demonstrated that the vector
coupling constant, $G_V$, has the same value for all thirteen transitions to within $\pm$0.013\%, thus confirming a key
part of the CVC hypothesis; and it sets an upper limit on a possible scalar current at 0.2\% of the vector current.  With
both these outcomes established, the results could then be used to extract a value for $V_{ud}$, the up-down
element of the CKM matrix, with which the top-row unitarity test of that matrix yielded the result 
0.9999(6) \cite{Towner2010a}.  This is in remarkable agreement with the Standard Model, and the tight uncertainty significantly limits the
scope for any new physics beyond the model.  Further tightening of the uncertainty would, of course, increase the impact
of this result even more.

Neglecting for now the possibility of any scalar current, we can relate the $ft$ value for a superallowed
$0^+$$\rightarrow$\,$0^+$ transition directly to the vector coupling constant, $G_V$ by the following
equation \cite{Hardy2009}:
\begin{equation}
\F t \equiv ft (1 + \delta_R^{\prime}) (1 + \delta_{NS} - \delta_C ) = \frac{K}{2 G_V^2 
(1 + \Delta^R_V )}~,
\label{Ftconst}
\end{equation}
where $\F t$ is defined to be the ``corrected" $ft$ value and $K/(\hbar c )^6 = 2 \pi^3 \hbar \ln 2/(m_e c^2)^5
= 8120.2787(11) \times 10^{-10}$ GeV$^{-4}$s.  There are four small correction terms: $\delta_C$ is the
isospin-symmetry-breaking correction; $\Delta^R_V$ is the transition-independent part of the radiative correction; and
the terms $\delta_R^{\prime}$ and $\delta_{NS}$ comprise the transition-dependent part of the radiative correction, 
the former being a function only of the maximum positron energy and the atomic number, $Z$, of the daughter nucleus, 
while the latter, like $\delta_C$, depends in its evaluation on the details of nuclear structure.  The two
structure-dependent terms $\delta_C$ and $\delta_{NS}$, which appear in Eq.~\ref{Ftconst} as a difference, together
contribute $\leq$1\% to most $\F t$ values \cite{Towner2008}.  Even so, at the current level of experimental precision, their
theoretical uncertainties contribute significantly to the final $\F t$-value uncertainties.

Experiments can help to reduce these theoretical uncertainties.  A method has recently been proposed \cite{Towner2010b},
by which the structure-dependent corrections can be validated.  The calculated corrections change considerably from
transition to transition, and the validation entails a comparison of these changes against the experimental changes
from transition to transition in the uncorrected $ft$ values.  In essence, validation depends on whether the calculated
corrections produce a result consistent with CVC.  The effectiveness of this validation process depends directly on the
experimental precision of the $ft$ values.  

The $ft$ value that characterizes any $\beta$-transition depends on three measured quantities: the total transition
energy, \qec; the half-life, $t_{1/2}$, of the parent state; and the branching ratio, $R$, for the particular
transition of interest.  The \qec-value is required to determine the statistical rate function, $f$, while the
half-life and branching ratio combine to yield the partial half-life, $t$.  It is important to recognize, though, that
$f$ varies approximately with the fifth power of \qec{}: If the fractional uncertainty in the measured \qec{}~value is
1$\times$$10^{-4}$, the corresponding uncertainty in $f$ is $\sim$5$\times$$10^{-4}$.  Thus, the precision required for
\qec-value measurements is substantially higher than that required for half-lives and branching ratios.

We report here \qec-value results for \nuc{10}{C}, \nuc{34}{Ar}, \nuc{38}{Ca} and \nuc{46}{V} with fractional
uncertainties in the range (1-5)$\times$$10^{-5}$, substantially better than any previous measurements for these
transitions, and low enough that, with improvements in their half-lives and branching ratios, the uncertainties in these
$\F t$ values could in principal be reduced to $\sim$$1\times10^{-4}$, a factor of five to ten below the uncertainties
of the best-known cases today.

The superallowed decay of \nuc{10}{C} is a particularly interesting case.  If scalar currents exist, they would lead
to a discrepancy between the $\F t$ values for the transitions in light nuclei and the average $\F t$ value for
the heavier nuclei (see Fig.~7 in Ref.~\cite{Hardy2009}).  In particular, the decays of \nuc{10}{C} and \nuc{14}{O} 
are the most sensitive to the presence of a scalar current.  Improved experimental precision for these two cases would
have a significant impact on the search for a scalar current.  If it were to be found, of course, that would constitute
new physics beyond the standard model.  Our \nuc{10}{C} measurement reported here is the first step along this path.

\section{EXPERIMENTAL METHOD}

\begin{table}[b]
 \caption{\label{tab:targets}The proton beam energies and target combinations used in these measurements.
Where applicable, the percentage of isotopic enrichment is given in parenthesis. Only the $^{46}$Ti target
was self supporting; all others were evaporated onto thin nickel foil. In all cases, the target thickness
was a few mg/cm$^2$. }
 \begin{ruledtabular}
 \begin{tabular}{llll}
  Target & $E_\textrm{protons}$        & Reaction & Product(s) \\ \hline \\[-1em]
 $^{10}$B ($\approx 90\%$)& 12 MeV   & $^{10}$B($p$,$n$)& $^{10}$C \\
                &     & $^{10}$B($p$,$p$)& $^{10}$B \\ \hline \\[-1em]
  KCl    & 35 MeV     & $^{35}$Cl($p$,2$n$)&$^{34}$Ar \\
         &            & $^{35}$Cl($p$,$pn$)&$^{34}$Cl+$^{34}$Cl$^m$ \\
         &            & $^{35}$Cl($p$,2$p$)&$^{34}$S \\ \hline \\[-1em]
  KCl    & 35 MeV     & $^{39}$K($p$,2$n$)&$^{38}$Ca \\
         &            & $^{39}$K($p$,$pn$)&$^{38}$K+$^{38}$K$^m$ \\
         &            & $^{39}$K($p$,2$p$)&$^{38}$Ar \\ \hline \\[-1em]
$^{46}$Ti ($>90\%)$ & 20 MeV & $^{46}$Ti($p$,$n$)&$^{46}$V \\
 \end{tabular}
 \end{ruledtabular}
\end{table}

The targets, proton beam energies, and reactions we employed in these measurements are listed in
Table~\ref{tab:targets}.

The experiments were carried out with the JYFLTRAP Penning-trap mass spectrometer at the University of
Jyv\"askyl\"a, Finland \cite{Jokinen2006}. The ions of interest were produced from fusion-evaporation
reactions induced by protons from the K130 cyclotron, with the reaction products being collected and
separated by the IGISOL technique \cite{Aysto2001}, which is both universal and fast, enabling
extraction of beams of any element within less than 100~ms.  The recoiling nuclei are primarily
slowed down in the target itself but are ultimately thermalized in a helium-filled stopping volume
\cite{Huikari2004}. The ions flow with helium out from the gas cell and into a sextupole ion guide
\cite{Karvonen2008}, after which they are electrostatically accelerated to an energy of 30$q$ keV. These energetic ions
are then separated with a 55$^\circ$ dipole magnet, which has a mass resolving power $R$ ($\equiv M/\Delta M$) of
about 500, and injected into a radiofrequency quadrupole (RFQ) structure for ion-beam cooling and
bunching \cite{Nieminen2001}.  Finally, each bunch is released to the JYFLTRAP Penning-trap setup where
the ions' masses are measured with the time-of-flight ion-cyclotron-resonance (TOF-ICR) technique
\cite{Graff1980}.

\subsection{Ion preparation}

\begin{figure}[t]
 \includegraphics[width=\columnwidth]{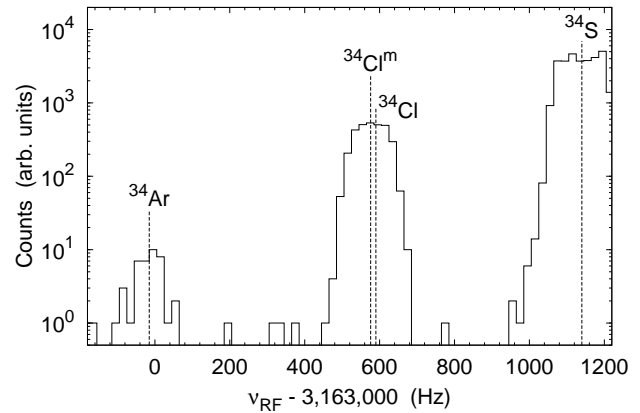}
 \caption{\label{fig:mass34_trap1}Quadrupole frequency scan of the purification trap. The trap was
tuned in this case to have a mass resolving power $R$ of about 30,000, which is enough
to separate isobars but not the isomeric states of $^{34}$Cl.}
\end{figure}

Ideally only one ion at a time is needed for a measurement, but in practice a few ions are usually used.
However, the ions of interest typically comprise less than 1\% of the mass-separated beam from IGISOL, so
to have, for example, a few $^{34}$Ar ions in a
bunch, we have to collect 2-3 orders-of-magnitude more ions --- mostly $^{34}$Cl and $^{34}$S --- in the RFQ
buncher.  Once a large enough bunch has been collected, it is sent to the first of the two Penning traps that
comprise the JYFLTRAP setup.  This first trap contains helium buffer gas and serves to purify the sample. In
it, the ions of interest are spatially separated with the sideband cooling technique \cite{Savard1991}.  After
separation, the ions are extracted towards the second Penning trap, their path to that trap being via an
electrode, in which there is a narrow central channel 2-mm in diameter. Only the centered ions of interest
can pass through this channel, while the other ions hit the electrode. The transmitted ions are then captured
in the second, precision Penning trap, which is operated in vacuum. There, the TOF-ICR mass measurement
could in principal be initiated.

\begin{figure}[t]
 \includegraphics[width=\columnwidth]{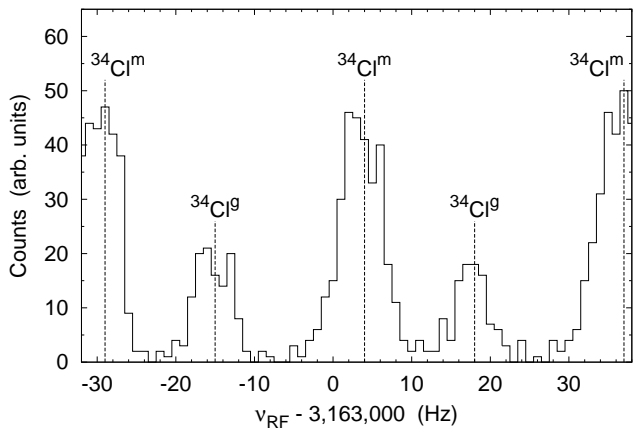}
 \caption{\label{fig:cl34_rclean}Dipole frequency scan in the precision trap for bunches containing both
$^{34}$Cl and $^{34}$Cl$^m$.  With an excitation time-pattern of 10/20/10~ms (on/off/on) a mass resolving power
$R\approx5\times 10^5$ was obtained. The two states of $^{34}$Cl are cleanly separated.}
\end{figure}

However, in the case of close-lying isomeric states purity is not yet assured.  As shown in
Fig.~\ref{fig:mass34_trap1} for the mass-34 measurements, the purification process in the first Penning trap
is sufficent to make clean bunches of $^{34}$Ar, but it is not enough to separate $^{34}$Cl from $^{34}$Cl$^m$.  
The same problem occurs in the case of mass-38 as well.  For these measurements we used the so-called Ramsey
cleaning technique \cite{Eronen2008a}, in which a further purification is accomplished by use of a dipole rf
electric field to drive the unwanted ions to large cyclotron orbits in the gas-free precision trap. The
excitation pattern and duration are chosen so that, upon completion, ions in the unwanted state have a large
orbit.  The ions are then transferred back to the purification trap and, en route, the unwanted ions hit
the electrode rather than passing through the narrow central channel.  An example of a cleaning frequency
scan is shown in Fig.~\ref{fig:cl34_rclean}.

\begin{figure}[b]
 \includegraphics[width=\columnwidth]{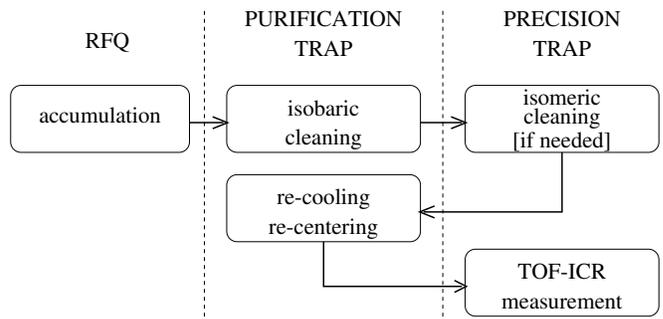}
 \caption{\label{fig:ions_go}The measurement cycle.}
\end{figure}

Even with ions that did not require such high-precision cleaning, we chose to transfer them back from
the precision trap to the purification trap. There the mono-isomeric ion sample was recooled and recentered with
the sideband cooling technique \cite{Savard1991}.  Only after this second purification step was the ion bunch
sent to the precision trap for the TOF-ICR mass measurement.  We found that this additional cooling step
significantly improved the quality of the measured TOF-ICR resonances and thus improved our
precision. The full ion preparation cycle is illustrated in Fig.~\ref{fig:ions_go}.

\subsection{TOF-ICR measurement}

In these measurements the time-of-flight ion-cyclotron resonance (TOF-ICR) technique \cite{Graff1980}
has been applied not only in conventional square-wave mode \cite{Konig1995} but also in
time-separated (Ramsey-type) mode \cite{George2007a,Kretzschmar2007}. The measurement procedure starts
with a phase-locked dipole rf electric field used to increase slightly the magnetron orbit radius of the
ions \cite{Blaum2003}. In measurements reported in this work, this excitation was applied for a single
magnetron period of about 5.5~ms and with an amplitude of about 50~mV.

Following the dipole magnetron excitation, a quadrupole excitation was switched on to couple the two
radial trap eigenmotions.  The frequency of this quadrupole excitation was scanned over a range that
included the sum of the two eigenfrequencies $\nu_c$, which is also the cyclotron frequency of ions
in the absence of any trapping electric fields: viz.
\begin{equation} \label{eq:qbm}
 \nu_c = \nu_+ + \nu_- = \frac{1}{2\pi} \frac{q}{m}B,
\end{equation}
where $\nu_+$ and $\nu_-$ are the frequencies of the two radial motions, commonly called the trap-modified
cyclotron and magnetron frequencies, respectively; $q/m$ is the charge-to-mass ratio of the ions and $B$ is
the magnetic field.  The duration we could use for the excitation was limited not only by the half-life of
the ions of interest, but also by ion-motion damping effects caused by residual gas present in the precision
trap.  These effects are much stronger with lighter ions. The excitation durations we used were between 
200~ms and 400~ms.

\begin{figure}
 \includegraphics[width=\columnwidth]{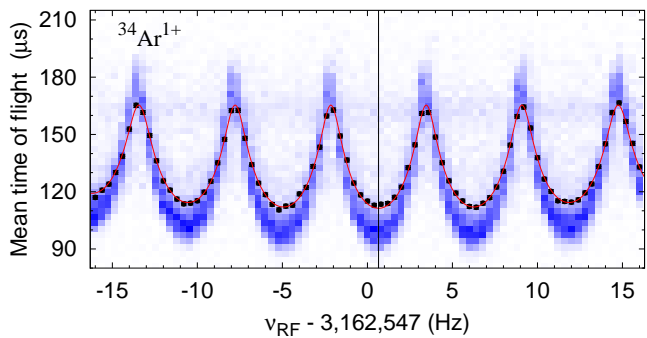}
 \caption{\label{fig:example_reso_ar34} (color online) A TOF-ICR curve measured for $^{34}$Ar$^+$ ions. An
excitation time pattern of 25/150/25~ms (on/off/on) was used. The black circles (with error bars smaller than
the points) are time-of-flight averages for each frequency. The (blue) pixels represent the number of detected
ions: the darker the pixel the more ions it represents. The solid (red) line is the fit to the experimental
data.  Note that the averages include some non-resonant background as well as the resonant $^{34}$Ar$^+$ ions,
which accounts for why the averages do not go through the densest concentration of pixels. See
Sec.~\ref{sec:ar34} for more details.}
\end{figure}

After completion of the quadrupole excitation, the ions were released from the trap in the direction of a
microchannel plate (MCP) detector.  As ions travel through a region with high magnetic field gradient, their
radial energy converts to axial velocity, so ions starting with more radial energy will gain more speed and
thus arrive earlier at the detector than the ions that have not been resonantly excited. Since the energy
content of the trap-modified cyclotron motion ($\nu_+$) is of the order of several eV and the energy of the
magnetron motion ($\nu_-$) is only a few $\mu$eV, the resonantly excited ions can have as much as a factor of
two shorter time-of-flight to the MCP detector than non-excited ions. Figure~\ref{fig:example_reso_ar34} shows
a sample TOF-ICR curve for $^{34}$Ar, in which the measured ion time-of-flight is plotted as a function of
the quadrupole-excitation frequency over a 30 Hz range.  In this case, the ion-motion excitation was accomplished
by use of Ramsey's method of time-separated oscillatory fields.

Figure \ref{fig:example_reso_ar34} also shows a fit to the experimental data.  For the mass-34 measurement
illustrated in the figure, and also for the mass-38 and mass-46 measurements, we took the shape of the Ramsey-type
resonance from Ref.~\cite{George2007a}.  For the mass-10 measurements we could not use the Ramsey procedure
(see Sec.~\ref{sec:10c} for more details) and
had to revert to the conventional resonance procedure, for which we took the shape of the resonance curve from
Ref.~\cite{Konig1995}. Very recently, the effects of ion-motion damping due to collisions with rest gas atoms
have been incorporated into the function describing the Ramsey resonance shape \cite{George2010}. Part of our
data was checked with both fitting functions. No significant shifts were seen in the results so, for the analysis
presented here, we used the function corresponding to the ideal lineshape.

\subsection{\qec{} value determination}

The \qec{} value is the total decay energy of the transition. It can be expressed
as the difference between the mass of the parent atom $M_\textrm{p}$ and that of the daughter $M_\textrm{d}$:
\begin{equation}
 Q_\textrm{EC} = \left( M_\textrm{p} - M_\textrm{d} \right ) c^2.
\end{equation}
In terms of the measured cyclotron frequencies for the singly-charged ions of the parent and daughter, 
$\nu_{c,\mathrm{p}}$ and $\nu_{c,\mathrm{d}}$ respectively [see Eq.~(\ref{eq:qbm})], the \qec{} value can be
written as
\begin{equation} \label{eq:QEC}
Q_\mathrm{EC} = \left ( \frac{\nu_{c,\mathrm{d}}}{\nu_{c,\mathrm{p}}} -1 \right ) \left (M_\mathrm{d} - m_e \right) + \Delta_{p,d},
\end{equation}
where $m_e$ is the electron rest mass and $\Delta_{p,d}$ arises from the atomic-electron binding-energy
difference between the parent and daughter atoms.  The latter contributes at most about 3~eV for the cases
we report on here, since we studied only singly-charged ions.

\subsection{Control of systematic errors}

The \qec{} values reported in this work were all measured as parent-daughter doublets, both with the
same value of $A/q$.  Thus we can apply Eq.~(\ref{eq:QEC}) to our results.  Furthermore, we obtained the
cyclotron frequency ratio by interleaving scans of the two ion species, about 30 s for one, then 30 s for
the other, with the alternation repeated for a number of hours.  The temporal drift of the magnetic field is
of the order of $3 \times 10^{-11}$~min$^{-1}$~\cite{Rahaman2007a}, so any shifts between one 30-s scan
and the next would have made a negligible contribution to the frequency ratio. 

We analyzed the data by splitting the alternating parent- and daughter-ion scans into approximately 30-minute
intervals, each consisting of about 30 pairs of scans.  The 30 scan-pairs were not merged to form just one
resonance curve for each ion species but several, the data being split according to the number of ions
recorded per bunch.  This allowed us to test for possible shifts in the resonance frequency due to multiple
ions being stored in the trap \cite{Kellerbauer2003}.  (Our procedures will be described more fully for each measured
\qec{} value in Sec.~\ref{res}.)  The time-of-flight resonance results obtained for each 30-minute interval
were then fitted separately for both ion species to get a frequency ratio.  The final frequency ratio for a
particular doublet was obtained from the weighted average of its interval results.

\begin{figure}[b]
 \includegraphics[width=\columnwidth]{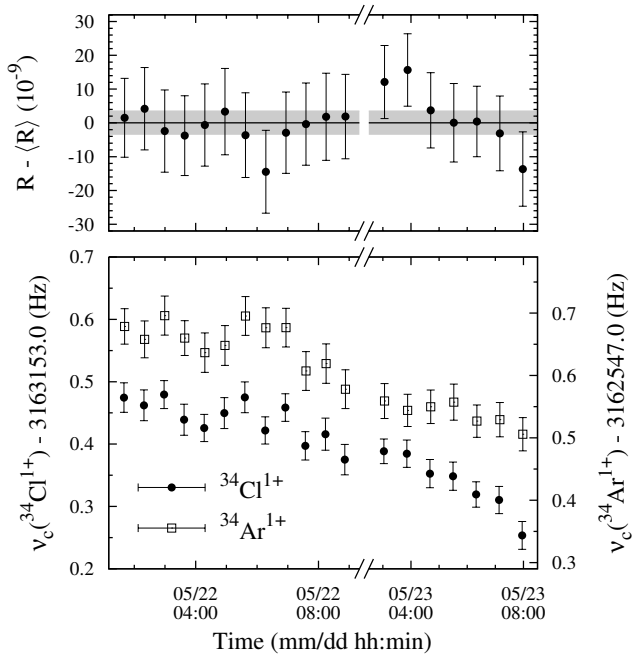}
 \caption{\label{fig:magfield_ar34} A series of fitted cyclotron frequencies for \nuc{34}{Ar} and \nuc{34}{Cl}
ions (lower panel).  Each frequency point includes the results of 30~interleaved scans. The top panel shows the
deviation of the corresponding frequency ratios, $\nu_d/\nu_m$ from the average frequency ratio.}
\end{figure}

As an example of the quality of results, Fig.~\ref{fig:magfield_ar34} shows the individual cyclotron
frequencies obtained for the \nuc{34}{Ar}-\nuc{34}{Cl} pair, together with the deviations of the frequency
ratios from the average value.  It can be seen that, although the magnetic field fluctuates, the
cyclotron-frequency ratios were consistent over an eight-hour period on one day, and a five-hour period
a day later. 

We have also considered other possible sources of systematic error.  Tiny differences between the measured
$\nu_+ + \nu_ -$ and the actual cyclotron frequency $\nu_c$ could result from a slight misalignment of the
electric- and magnetic-field axes, and from distortion in the quadrupole electric field \cite{Gabrielse2009}.
Because the ion-pairs in our measurements are $A/q$ doublets, the effect on the frequency ratio would be
negligibly small compared to the statistical uncertainty.  Mass-dependent shifts \cite{Elomaa2009b} are 
negligible as well, also because we are working with doublets having the same mass number.  Previous
measurements with JYFLTRAP have successfully reproduced accurately known isomeric excitation energies or
\qec{} values \cite{Rahaman2008a, Eronen2009a} down to a relative precision of $\Delta Q/M \approx 2\times
10^{-9}$.

In the measurements reported here, by far the largest contribution to each final uncertainty is the statistical
component, which originates from counting statistics and the fitting procedure.

\section{RESULTS AND ANALYSIS}
\label{res}
The $Q_\textrm{EC}$ values of the superallowed $\beta$ emitters $^{10}$C, $^{34}$Ar, $^{38}$Ca and $^{46}$V
were all measured during a 9-day period of beam time in May 2010, only a month before the IGISOL and JYFLTRAP
facilities were shut down in preparation for being moved to a different target location. Our four different
frequency-ratio measurements are described individually in the following sections.

\subsection{$^{10}$C}
\label{sec:10c}
The \qec{} value for $^{10}$C proved to be the most difficult one we have measured so far. Mass-10 being the
lightest mass ever measured with JYFLTRAP, we carefully tuned the setup before the on-line experiment began, using
stable $^{10}$B and $^{12}$C ions from an off-line ion source. We found that the buffer-gas pressure of the
purification Penning trap had to be significantly reduced since the cooling effect of the helium gas is much
stronger for light ions.  Even so, damping of the TOF-ICR resonance was rather pronounced and thus we could only
use short excitation times.  An additional difficulty was that the transmission of the RFQ was rather poor.
Nevertheless, in the end enough ions could be delivered to the Penning trap for a successful measurement.

Unlike our experience with heavier ions ($A$\,$\geq$\,$23$), a rather strong dipole component in the quadrupole field
was apparent, as evidenced by a clear resonance observed at frequency $\nu_+$, about 170~Hz away from the
$\nu_++\nu_-$ sideband.  This prevented our using Ramsey excitation, since the two resonance patterns
overlapped. Instead, we used a conventional TOF-ICR resonance technique with a 200~ms excitation time.

\begin{figure}[b]
 \includegraphics[width=\columnwidth]{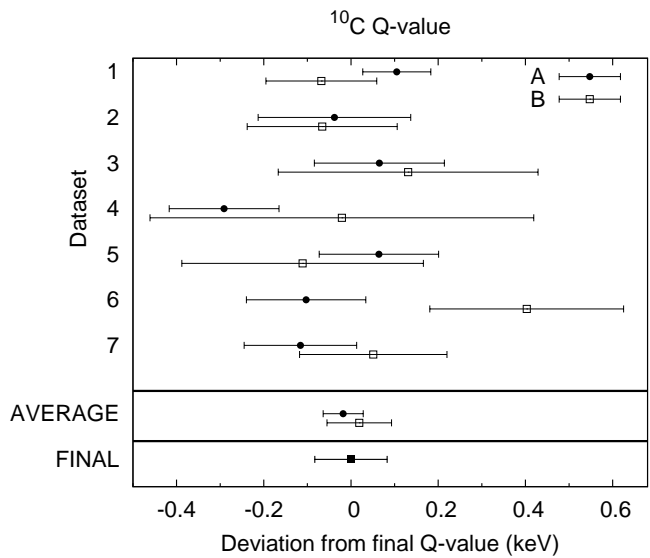}
 \caption{\label{fig:c10}\qec{} values obtained for each set of 30 scans. All the data were analyzed to account
for multiple stored ions. In the data denoted by ``A'' the maximum number of ions per accepted bunch was 5,
while those denoted ``B'' included a maximum of 3. We obtained the final value by taking the unweighted
average of ``A'' and ``B'' and applying an error bar derived by taking the larger uncertainty of the two and
adding the difference between ``A'' and ``B'' in quadrature.}
\end{figure}

Several checks of the data were done in order to ensure that the measurements yielded correct results despite the
strong dipole component.  In addition to measuring the \qec{} value between $^{10}$C and $^{10}$B, we also measured the well known
mass of stable $^{13}$C \cite{Audi2003}, using $^{12}$C as the reference ion.  Also, parameters like dipole-magnetron-excitation
amplitude and ion-transfer time between the two traps were varied to check that there was no shift in the
measured frequency ratio when the ions occupied a different fraction of the trap volume.

In all, we obtained seven sets of data for the $^{10}$C \qec{} value and three for the $^{13}$C mass.  The data
were analyzed using count-rate class analysis \cite{Kellerbauer2003} to account for the effects of multiple
stored ions in the trap.  Moreover, to double-check our results, the class division was done in two different ways.
In the first analysis, we subdivided the data into three classes according to how many ions were in each bunch, 
1-2, 3 or 4-5.  In the second analysis only two classes were retained, those with 1 ion/bunch and those with
2-3 ions/bunch.  As can be seen from Fig.~\ref{fig:c10}, the results from each group of 30 interleaved scans
changed from one analysis to the other but the two average frequency ratios were consistent with one another.
Our final frequency ratio and \qec{} value appear in Table~\ref{tab:results}, and the latter is compared with
previous measurements of the \qec{} value derived from ($p$,$n$) threshold measurements \cite{Barker1984,
Barker1998} in Fig.~\ref{fig:All_mass_comparison}.

The \qec{} value given in Table~\ref{tab:results} is, of course, for the ground-state-to-ground-state transition.
The superallowed transition feeds the $0^+$ state in $^{10}$B, which is at 1740.07(2)~keV \cite{Hardy2009}, so
our result for the transition populating the ground state corresponds to a \qec{} value for the superallowed transition
of 1908.05(8)~keV. 

\begin{table}[t]
 \caption{\label{tab:results} The obtained frequency ratios and the derived \qec{} values or
energy differences for ions having $A$=10, 34, 38 and 46.}
 \begin{ruledtabular}
 \begin{tabular}{llll}
  Ion A & Ion B & Frequency ratio $\frac{\nu_B}{\nu_A}$ & \qec{} or $\Delta E$ (keV) \\ \hline \\[-1em]
  Mass 10: & & & \\
  \nuc{10}{C} & \nuc{10}{B}  & 1.000 391 157(9)   &  \phantom{1}3648.12(8)  \\ [1 mm]
  Mass 34: & & & \\
  \nuc{34}{Ar} & \nuc{34}{Cl}  & 1.000 191 551 8(27)   &  \phantom{1}6061.82(9)  \\
  \nuc{34}{Ar} & \nucm{34}{Cl} & 1.000 186 923 2(33)   &  \phantom{1}5915.37(10)   \\
  \nucm{34}{Cl} & \nuc{34}{Cl} & 1.000 004 630(8)    &  \phantom{11}146.52(26)   \\
  \nuc{34}{Ar} & \nuc{34}{S}   & 1.000 365 151(6)    &  11553.51(19) \\ [1mm]
  Mass 38: & & & \\
  \nuc{38}{Ca} & \nucm{38}{K} & 1.000 186 955 0(28) & \phantom{1}6612.15(10) \\
  \nuc{38}{Ca} & \nuc{38}{K}  & 1.000 190 632 6(27) & \phantom{1}6742.19(10) \\
  \nucm{38}{K} & \nuc{38}{K}  & 1.000 003 681 5(38) & \phantom{11}130.21(14) \\
  \nuc{38}{Ca} & \nuc{38}{Ar}& 1.000 357 913 2(32) & 12656.36(11) \\ [1mm]
  Mass 46: & & & \\
  \nuc{46}{V} & \nuc{46}{Ti}  & 1.000 164 760 8(23)   &  \phantom{1}7052.44(10)  \\
 \end{tabular}
 \end{ruledtabular}
\end{table}

Our control measurement of the $^{13}$C-to-$^{12}$C mass difference yielded a frequency ratio of
$\nu_c(^{12}$C$)/\nu_c(^{13}$C$)$\,\,=\,\,1.083\,\,616\,\,728(5).  This
corresponds to a mass excess for $^{13}$C of 3125.04(6)~keV, which is in perfect agreement with the high-precision
literature value of 3125.011(1)~keV \cite{Audi2003}.  This further confirms that our \nuc{10}{C} \qec{} value does
not suffer from any significant systematic error.

\subsection{\label{sec:ar34} $^{34}$Ar}

To determine the \qec{} value for $^{34}$Ar, we measured its frequency ratio, not only compared with $^{34}$Cl but
also with the high-spin ($3^+$) isomer $^{34}$Cl$^{m}$ and with its grand-daughter $^{34}$S.  The two states in
$^{34}$Cl are only 146 keV apart, but were easily separated as we have done before \cite{Eronen2009a} using the
Ramsey cleaning method.  The cyclotron frequencies were measured using a Ramsey-type excitation with the pattern
25/150/25~ms (on/off/on).  This short excitation time was used because of the residual-gas impurities
present in the precision trap.  When we used longer times, we observed significant charge-exchange losses
for all ion species. The effect of these losses is also evident in Fig.~\ref{fig:example_reso_ar34}, where
some ions are seen to appear at $\approx$170 $\mu$s time-of-flight regardless of the excitation frequency.
This accounts for why the average time-of-flight points do not go through the darkest concentration of pixels
in the figure: the contaminants, which appear at $\approx$170 $\mu$s and are probably singly charged O$_2$
molecules, serve to pull the average up a bit.

To take this constant time-of-flight background into account, we added two additional fit parameters: a constant
time of flight for the background, and the ratio of the number of background counts to that of the resonant ions.
Despite the presence of contaminating ions, no systematic shifts of the fitted frequencies were observed as we
increased the number of stored ions in the trap.

\begin{figure}[t]
 \includegraphics[width=\columnwidth]{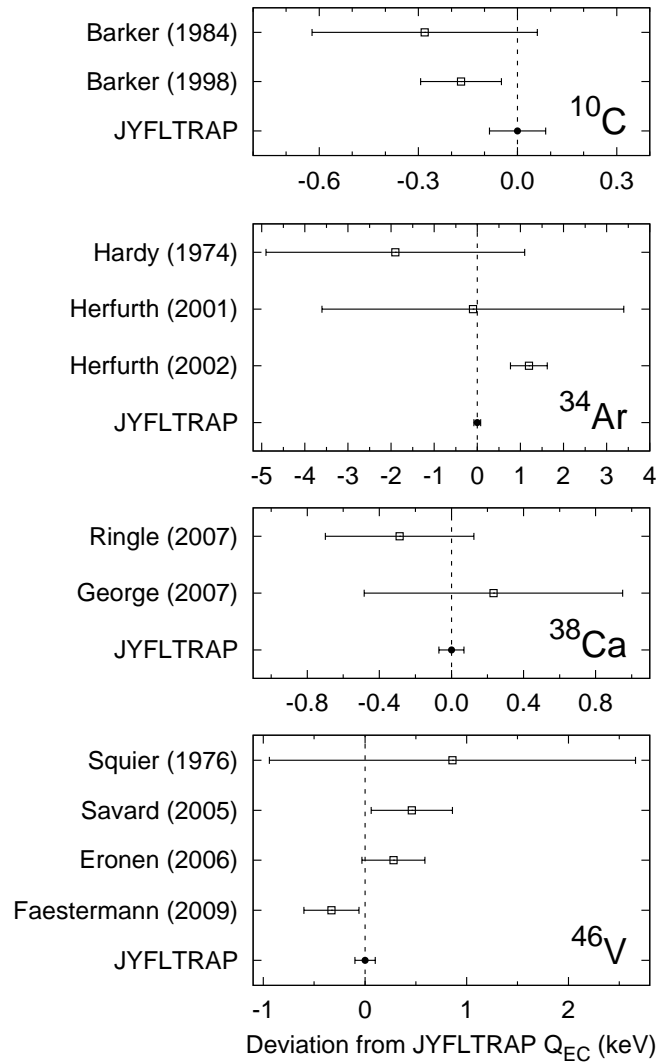}
 \caption{\label{fig:All_mass_comparison}Comparison of our \qec{} value measurements, labelled JYFLTRAP,
with previous measurements.  For each transition, our result is plotted at 0 on the abscissa, and
the other results are plotted as differences $\left ( Q_\mathrm{EC}^\mathrm{LIT.} - Q_\mathrm{EC}^\mathrm{JYFLTRAP}\right)$.
The references from the top down are \cite{Barker1984,
Barker1998, Hardy1974b, Herfurth2001, Herfurth2002, Ringle2007, George2007, Squier1976, Savard2005,
Eronen2006, Faestermann2009} respectively.}
\end{figure}

\begin{table}[b]
 \caption{\label{tab:ar34_finals}\qec{} values for the $^{34}$Ar-to-$^{34}$Cl superallowed transition obtained
 via three different reference ions.  The input data are taken from Table \ref{tab:results} and from
 Refs.~\cite{Snover1971, Eronen2009a}.}
 \begin{ruledtabular}
 \begin{tabular}{ll}
  Method & \qec{} (keV) \\ \hline \\[-1em]
  \nuc{34}{Ar}---\nuc{34}{Cl} direct  & 6061.82(9)  \\
  via \nucm{34}{Cl}  & 6061.89(28) \\
  via \nuc{34}{S}    & 6061.85(20)\footnotemark[1] \\ [1mm]
  FINAL & \textbf{6061.83(8)} \\
  \end{tabular}
 \end{ruledtabular}
\footnotetext[1]{The mass difference between \nuc{34}{Cl} and \nuc{34}{S} [5491.662(47)~keV] was taken from
Ref.~\cite{Eronen2009a}.}
\end{table}

The frequency ratios and \qec{} values we obtained from these measurements are compiled in Table \ref{tab:results}
and the final derived \qec{} values are in Table \ref{tab:ar34_finals}.  Our value for the excitation energy of
the $^{34}$Cl isomer, 146.52(26)~keV, agrees well with our previous JYFLTRAP value, 146.29(10)~keV, which we reported
in 2009 \cite{Eronen2009a}, and with a more precise value, 146.36(3)~keV, obtained \cite{Snover1971} from its decay
$\gamma$ ray.  Furthermore, the three \nuc{34}{Ar}-\nuc{34}{Cl} \qec{} values, derived via three different
paths, are in remarkable agreement with one another.  Our final result for the \qec{} value of this superallowed
transition is 6061.83(8)~keV.

This value is compared with previous measurements of the \nuc{34}{Ar} \qec{} value \cite{Hardy1974b, Herfurth2001,
Herfurth2002} in Fig.~\ref{fig:All_mass_comparison}.  The earliest of these results \cite{Hardy1974b} was from a
($p$,$n$) $Q$-value measurement; the later two \cite{Herfurth2001, Herfurth2002} were Penning-trap measurements
from ISOLTRAP.   Only one of these results---the most recent value from ISOLTRAP\cite{Herfurth2002}---has an uncertainty
comparable to ours, although even its uncertainty is five times larger than ours.  However, our result disagrees
by three of the latter's standard deviations.
We have no definitive explanation for this discrepancy but there is an important difference between the two
measurements: ours obtained the \nuc{34}{Ar} \qec{} value directly by a measurement of the frequency ratio of the
daughter to the parent ions.  The ISOLTRAP measurement used
\nuc{39}{K} as its reference ion.  Thus, to get the \nuc{34}{Ar} \qec{} value, the mass of the daughter
\nuc{34}{Cl} also had to be linked to \nuc{39}{K}.  This link via \nuc{39}{K} -- 5 mass units away -- may well
have been the source of error.

%We have no definitive explanation for this discrepancy
%but there is an important difference between the two measurements: ours obtained the \nuc{34}{Ar} \qec{} value
%directly by a measurement of the frequency ratio of the daughter to the parent ions.  The ISOLTRAP measurement used
%\nuc{39}{K} as its reference ion.  Thus, to get the \nuc{34}{Ar} \qec{} value, the mass of the daughter
%\nuc{34}{Cl} also had to be linked to \nuc{39}{K}.  It is possible that this essential link, as it appears in the
%current mass tables \cite{Audi2003}, is not correct.  

\subsection{\nuc{38}{Ca}}
Our measurement of the \nuc{38}{Ca} \qec{} value was conducted in the same way as the measurement just described for
\nuc{34}{Ar}.  The daughter nucleus in this case, \nuc{38}{K}, has a low-lying isomeric state just like
\nuc{34}{Cl} has, although it is the isomeric state in \nuc{38}{K} that has spin and parity of $0^+$ and the ground
state that is $3^+$.  These states are only 130 keV apart but were easily separated with the Ramsey cleaning method.  
As with the \nuc{34}{Ar} measurement, we obtained the \qec{} value for \nuc{38}{Ca}, not only by a direct
daughter-parent frequency ratio, but also via the high-spin ground state of \nuc{38}{K} and via the granddaughter
nucleus \nuc{38}{Ar}. Charge-exchange background was evident in the \nuc{38}{Ar} frequency measurement but not for
the other ion species. We used a Ramsey excitation pattern of 25/350/25~ms (on/off/on) for all measurements except
for one set of data connecting \nuc{38}{Ca} to \nuc{38}{Ar}.  In that case, we used a shorter excitation pattern, 
25/150/25~ms, in order to confirm that there was no change in the results as the number of contaminant ions increased.

Our measured frequency ratios and \qec{} values appear in Table~\ref{tab:results} and the final derived \qec{} values
are in Table \ref{tab:ca38_finals}.  Our measured value for the excitation energy of the \nuc{38}{K} isomer,
130.21(14)~keV, agrees well with our previous JYFLTRAP value, 130.13(6) keV, and with the previously accepted
value of 130.4(3)~keV
\cite{Endt1990}, which is the least precise.  Furthermore, the three \nuc{38}{Ca}-\nucm{38}{K} \qec{} values, derived
via three different paths, are in excellent agreement with one another.  Our final result for the \qec{} value of this
superallowed transition is 6612.12(7)~keV.

\begin{table}
 \caption{\label{tab:ca38_finals} \qec{} values for the $^{38}$Ca-to-\nucm{38}{K} superallowed transition obtained
via three different reference ions. The input data are taken from Table \ref{tab:results}
and from Ref.~\cite{Eronen2009a}.}
 \begin{ruledtabular}
 \begin{tabular}{ll}
  Method & \qec{} (keV) \\ \hline \\[-1em]
  \nuc{38}{Ca}---\nucm{38}{K} direct  & 6612.15(10)  \\
  via \nuc{38}{K}     & 6611.99(16) \\
  via \nuc{38}{Ar}    & 6612.14(12)\footnotemark[1] \\[1mm]
  FINAL & \textbf{6612.12(7)} \\
  \end{tabular}
 \end{ruledtabular}
\footnotetext[1]{The mass difference between \nucm{38}{K} and \nuc{38}{Ar} [6044.223(41)~keV]
was taken from Ref.~\cite{Eronen2009a}.}
\end{table}

This value is compared in Fig.~\ref{fig:All_mass_comparison} with the two previous determinations of the \nuc{38}{Ca}
\qec{} value, both based on Penning-trap mass measurements, one from the LEBIT trap \cite{Ringle2007} and the
other from ISOLTRAP \cite{George2007}.  These two values, which are quite recent, and the new measured result we
report here all agree within error bars. Our value, though, is about six times more precise than that in 
Ref.~\cite{Ringle2007} and ten times more precise than Ref.~\cite{George2007}.

\subsection{\nuc{46}{V}}

We have already measured the \qec{} value for the superallowed decay of \nuc{46}{V} once before at JYFLTRAP, in 2006
\cite{Eronen2006}. Our motivation for remeasuring it now was to improve the precision of the result.  Since 2006 we have
introduced a number of improvements to our system, most notably the rapid alternation of parent-daughter frequency
scans and the use of Ramsey excitation.   The excitation pattern used for this measurement was 25/350/25~ms
(on/off/on).  The resonances we obtained were clean and showed no evidence of any charge-exchange products.  The frequency
ratio and corresponding \qec{} value are presented in Table~\ref{tab:results}.

Our new \qec{} value is compared with the four previous measurements of the \nuc{46}{V} \qec{} value \cite{Squier1976,
Savard2005, Eronen2006, Faestermann2009} in Fig.~\ref{fig:All_mass_comparison}.  Our new result agrees with our previous
one \cite{Eronen2006} but has an uncertainty smaller by a factor of three.  In fact, our new result also agrees with the
other three measurements, one of which is from the CPT Penning trap \cite{Savard2005}, another is from a (\nuc{3}{He},$t$) 
reaction $Q$ value \cite{Faestermann2009}, and the third is from a ($p$,$n$) threshold measurement \cite{Squier1976}.
All three have uncertainties at least a factor of three greater than our new result.

\section{Conclusions}

Our four new \qec{}-value results for superallowed transitions are collected in Table~\ref{tab:all_results}, where they
are compared with the equivalent values that appeared in the most recent survey of superallowed $0^+$$\rightarrow 0^+$
nuclear $\beta$ decay \cite{Hardy2009}.  In all cases, our new results have reduced the uncertainties considerably,
although in the case of \nuc{34}{Ar} the reduction is constrained by the inconsistency between our result and one of
the previous measurements \cite{Herfurth2002} (see Fig.~\ref{fig:All_mass_comparison}).  That inconsistency leads to a
normalized $\chi^2$ of 7 for the average and, following the procedures used in Ref.~\cite{Hardy2009}, we increase the
uncertainty on the average by a scale factor equal to the square root of the normalized $\chi^2$.

\begin{table}[t]
 \caption{\label{tab:all_results}The four \qec{} values for superallowed transitions that were obtained in this
work.  Also shown are the equivalent values quoted in the most recent survey of data \cite{Hardy2009} and the
new weighted averages including our measurements.}
\begin{ruledtabular}
\begin{tabular}{llccc}
           &                  &  \multicolumn{3}{c}{\qec{} values (keV)}  \\[1mm]
\cline{3-5} \\[-2mm]
 Parent    & Daughter         &  this work &  survey\cite{Hardy2009}  &  average  \\[1mm]
 \hline \\[-2mm]

 $^{10}$C  & $^{10}$B($0^+$)  & 1908.05(8) & 1907.87(11) & 1907.99(7) \\
 $^{34}$Ar & $^{34}$Cl        & 6061.83(8) & 6062.98(48) & ~6061.86(21) \\
 $^{38}$Ca & $^{38}$K$^m$     & 6612.12(7) & 6611.75(41) & 6612.11(7) \\
 $^{46}$V  & $^{46}$Ti        & ~7052.44(10)& 7052.40(16) & 7052.45(9) \\
\end{tabular}
\end{ruledtabular}
\end{table}

Although our improvement in \qec{}-value precision for these four cases is significant, our results do not in
themselves reduce the uncertainty in the corresponding $\F t$ values.  For each case, the uncertainty in its $\F t$
value is dominated by another property of the transition: For \nuc{10}{C}, \nuc{34}{Ar} and \nuc{38}{Ca}, it is the
branching ratio that dominates, while for \nuc{46}{V} it is the half-life.  What our results do is
to provide \qec{} values with fractional uncertainties that are comfortably below what is likely to be achieved in
the near future for branching ratios or half-lives.  Thus, whatever experimental improvements can be achieved in
reducing the branching-ratio uncertainties for the decays of \nuc{10}{C}, \nuc{34}{Ar} and \nuc{38}{Ca}, that
reduction will translate directly into reduced $\F t$-value uncertainties; and the same argument applies to the
half-life of \nuc{46}{V}.

To give one example, the branching ratio for the superallowed transition from \nuc{34}{Ar} is currently known to
a fractional uncertainty of $2.6\times10^{-3}$ and its half-life to $4.7\times10^{-4}$ \cite{Hardy2009}.  The
fractional uncertainty we report here for its \qec{} value is $1.3\times10^{-5}$, which corresponds to a
fractional uncertainty on the statistical rate function $f$ of $7.4\times10^{-5}$, a factor of six better than
the half-life and a factor of 35 better than the branching ratio.  Because \nuc{34}{Ar} has a rather favorable
decay scheme, it should be possible with currently available techniques to reduce the branching-ratio uncertainty
to $1.0\times10^{-3}$ or even below that.  This would lead to an $\F t$-value for \nuc{34}{Ar} at essentially that
same precision.  Then it would become possible for the first time to compare at the 0.1\% level a mirror pair of
superallowed transitions, \nuc{34}{Ar}$\rightarrow$\nuc{34}{Cl} and \nuc{34}{Cl}$\rightarrow$\nuc{34}{S}, a
comparison that would help to distinguish among the various models used to calculate the isospin-symmetry-breaking
correction to superallowed decays \cite{Towner2010b}.

It is also interesting to note that our measurement has slightly increased the \qec{} value for the \nuc{10}{C}
superallowed decay.  If we take the average \qec{} value listed in Table~\ref{tab:all_results} and include a new half-life
measurement for \nuc{10}{C} \cite{Barker2009} together with the data listed in the 2009 survey \cite{Hardy2009}, we
obtain an $\F t$ value for the \nuc{10}{C} superallowed transition of $3077.9(45)$ s.  This is slightly outside
error bars from the average of all $\F t$ values, 3072.08(79) s, obtained in the 2009 survey.  If this discrepancy
were to be confirmed by an improved branching-ratio value for \nuc{10}{C} and by a similarly high $\F t$ value for
the \nuc{14}{O} decay, it could signal the appearance of a scalar current (see Ref.~\cite{Hardy2009}).  With this
motivation, it is our plan in future to measure the \qec{}-value of the superallowed transition from \nuc{14}{O}.  
Obviously, an improved value for the \nuc{10}{C} branching ratio would also be very welcome.

\begin{acknowledgments}

This work has been supported by the EU 6th Frame-
work programme ``Integrating Infrastructure Initiative-
Transnational Access'' Contract No. 506065 (EURONS) and
by the Academy of Finland under the Finnish Centre of
Excellence Programme 2006-2011 (Nuclear and Accelerator
Based Physics Programme at JYFL). JCH was supported by the U.\,S. Department
of Energy under Grant DE-FG03-93ER40773 and by the Robert A. Welch
Foundation under Grant no. A-1397.
\end{acknowledgments}

%\bibliography{/home/tomero/articles/bibtexiin/tommi_bib_daa.bib}

\begin{thebibliography}{10}

\bibitem{Towner2010a}
I.S. Towner and J.C. Hardy, Rep. Prog. Phys. {\bf 73}, 046301 (2010).

\bibitem{Hardy2009}
J.~C. Hardy and I.~S. Towner, Phys. Rev. C {\bf 79},  055502  (2009).

\bibitem{Towner2008}
I.S. Towner and J.C. Hardy, Phys. Rev. C {\bf77}, 025501 (2008).

\bibitem{Towner2010b}
I.S. Towner and J.C. Hardy, Phys. Rev. C {\bf 82}, 065501 (2010)

\bibitem{Jokinen2006}
A. Jokinen {\it et~al.}, Int. J. Mass Spectrom. {\bf 251},  204  (2006).

\bibitem{Aysto2001}
J. \"Ayst\"o, Nucl. Phys. A {\bf 693},  477  (2001).

\bibitem{Huikari2004}
J. Huikari {\it et~al.}, Nucl. Instrum. Methods Phys. Res., Sect. B {\bf 222},
  632  (2004).

\bibitem{Karvonen2008}
P. Karvonen {\it et~al.}, Nucl. Instrum. Methods Phys. Res., Sect. B {\bf 266},
   4794  (2008).

\bibitem{Nieminen2001}
A. Nieminen {\it et~al.}, Nucl. Instrum. Methods Phys. Res., Sect. A {\bf 469},
   244  (2001).

\bibitem{Graff1980}
G. Gr{\"a}ff, H. Kalinowsky, and J. Traut, Z. Phys. A {\bf 297},  35  (1980).

\bibitem{Savard1991}
G. Savard {\it et~al.}, Phys. Lett. A {\bf 158},  247  (1991).

\bibitem{Eronen2008a}
T. Eronen {\it et~al.}, Nucl. Instrum. Methods Phys. Res., Sect. B {\bf 266},
  4527  (2008).

\bibitem{Konig1995}
M. K\"onig {\it et~al.}, Int. J. Mass Spectrom. {\bf 142},  95  (1995).

\bibitem{George2007a}
S. George {\it et~al.}, Int. J. Mass Spectrom. {\bf 264},  110  (2007).

\bibitem{Kretzschmar2007}
M. Kretzschmar, Int. J. Mass Spectrom. {\bf 264},  122  (2007).

\bibitem{Blaum2003}
K. Blaum {\it et~al.}, J. Phys. B: At., Mol. Opt. Phys. {\bf 36},  921  (2003).

\bibitem{George2010}
S. George {\it et~al.}, Int. J. Mass Spectrom. {\bf 299}, 102  (2010).

\bibitem{Rahaman2007a}
S. Rahaman {\it et~al.}, Eur. Phys. J. A {\bf 34},  5  (2007).

\bibitem{Kellerbauer2003}
A. Kellerbauer {\it et~al.}, Eur. Phys. J. D {\bf 22},  53  (2003).

\bibitem{Gabrielse2009}
G. Gabrielse, Int. J. Mass Spectrom. {\bf 279},  107  (2009).

\bibitem{Elomaa2009b}
V.-V. Elomaa {\it et~al.}, Nucl. Instrum. Methods Phys. Res., Sect. A {\bf
  612},  97  (2009).

\bibitem{Rahaman2008a}
S. Rahaman {\it et~al.}, Phys. Lett. B {\bf 662},  111  (2008).

\bibitem{Eronen2009a}
T. Eronen {\it et~al.}, Phys. Rev. Lett. {\bf 103},  252501  (2009).

\bibitem{Barker1988}
P.~H. Barker and S.~M. Ferguson, Phys. Rev. C {\bf 38},  1936  (1988).

\bibitem{Baker1989}
S.~C. Baker, M.~J. Brown, and P.~H. Barker, Phys. Rev. C {\bf 40},  940
  (1989).

\bibitem{Audi2003}
G. Audi {\it et~al.}, Nucl. Phys. {\bf A729}, 327 (2003).

\bibitem{Barker1984}
P.~H. Barker and R.~E. White, Phys. Rev. C {\bf 29},  1530  (1984).

\bibitem{Barker1998}
P.~H. Barker and P.~A. Amundsen, Phys. Rev. C {\bf 58},  2571  (1998).

\bibitem{Snover1971}
K.~A. Snover {\it et~al.}, Phys. Rev. C {\bf 4},  398  (1971).

\bibitem{Hardy1974b}
J.~C. Hardy {\it et~al.}, Phys. Rev. C {\bf 9},  252  (1974).

\bibitem{Herfurth2001}
F. Herfurth {\it et~al.}, Nucl. Instrum. Methods Phys. Res., Sect. A {\bf 469},
   254  (2001).

\bibitem{Herfurth2002}
F. Herfurth {\it et~al.}, Eur. Phys. J. A {\bf 15},  17  (2002).

\bibitem{Endt1990}
P.M. Endt, Nucl. Phys. A {\bf 521}, 1 (1990).

\bibitem{Ringle2007}
R. Ringle {\it et~al.}, Int. J. Mass Spectrom. {\bf 262},  33  (2007).

\bibitem{George2007}
S. George {\it et~al.}, Phys. Rev. Lett. {\bf 98},  162501  (2007).

\bibitem{Squier1976}
G.T.A. Squier, W.E. Burcham, S.D. Hoath, J.M. Freeman, P.H. Barker and R.J. Petty, Phys. Lett. {\bf 65B}, 122 (1976).

\bibitem{Savard2005}
G. Savard {\it et~al.}, Phys. Rev. Lett. {\bf 95}, 102501 (2005).

\bibitem{Eronen2006}
T. Eronen {\it et~al.}, Phys. Rev. Lett. {\bf 97}, 232501 (2006).

\bibitem{Faestermann2009}
T. Faestermann, R. Hertenberger, H.-F. Wirth, R. Kr\"{u}cken, M. Mahgoub and P. Maier-Komor, Eur. Phys. J. A,
{\bf 42}, 339 (2009).

\bibitem{Barker2009}
P.H. Barker, K.K.H. Leung and A.P. Byrne, Phys. Rev. C {\bf}79, 024311 (2009).
 

\end{thebibliography}

\end{document}